\documentclass[10pt,twocolumn,letterpaper]{article}

\usepackage{titlesec}
\titleformat*{\section}{\large\bfseries}
\titleformat*{\subsection}{\normalsize\bfseries}

\usepackage{cite}
\usepackage{amsmath,amssymb,amsfonts}
\usepackage{algorithmic_mod2}
\usepackage{graphicx}
\usepackage{textcomp}
\usepackage{wrapfig}
\usepackage{pifont}
\usepackage[dvipsnames]{xcolor}
\usepackage{url}
\usepackage{soul}

\usepackage{multirow}
\usepackage{array}
\newcolumntype{x}[1]{>{\centering\let\newline\\\arraybackslash\hspace{0pt}}m{#1}}
\usepackage{hyperref}

\usepackage{algorithm}

\usepackage{authblk}

\begin{document}
\title{Accurate Simulation Pipeline for Passive Single-Photon Imaging}
\author[1,2]{Aleksi Suonsivu\thanks{A. Suonsivu and L. Salmela contributed equally. This work is an extension of the conference workshop paper titled 'Time-Resolved MNIST Dataset for Single-Photon Recognition' presented at ECCV Synthetic Data for Computer Vision Workshop, 2024. Paper can be found here: \url{https://doi.org/10.1007/978-3-031-91907-7_8}.}}
\author[1]{Lauri Salmela}
\author[1]{Leevi Uosukainen}
\author[3]{Edoardo Peretti}
\author[1]{Radu Ciprian Bilcu}
\author[3]{Giacomo Boracchi}

\renewcommand\footnotemark{}


\affil[1]{\small Huawei Technologies Finland Oy, Tampere, Finland, \texttt{\{name.surname\}@huawei.com}}
\affil[2]{\small Tampere University, Tampere, Finland, \texttt{\{name.surname\}@tuni.fi}}
\affil[3]{\small Politecnico di Milano, Milano, Italy, \texttt{\{name.surname\}@polimi.it}}

\date{}

\maketitle

\begin{abstract}

Single-Photon Avalanche Diodes (SPADs) are new and promising imaging sensors. These sensors are sensitive enough to detect individual photons hitting each pixel, with extreme temporal resolution and without readout noise. Thus, SPADs stand out as an optimal choice for low-light imaging. Due to the high price and limited availability of SPAD sensors, the demand for an accurate data simulation pipeline is substantial. Indeed, the scarcity of SPAD datasets hinders the development of SPAD-specific processing algorithms and impedes the training of learning-based solutions. 

In this paper, we present a comprehensive SPAD simulation pipeline and validate it with multiple experiments using two recent commercial SPAD sensors. Our simulator is used to generate the SPAD-MNIST, a single-photon version of the seminal MNIST dataset, to investigate the effectiveness of convolutional neural network (CNN) classifiers on reconstructed fluxes, even at extremely low light conditions, e.g., 5~mlux. We also assess the performance of classifiers exclusively trained on simulated data on real images acquired from SPAD sensors at different light conditions. The synthetic dataset encompasses different SPAD imaging modalities and is made available for download. Project page: \url{https://boracchi.faculty.polimi.it/Projects/SPAD-MNIST.html}.

\end{abstract}

\section{Introduction}

Pixels in traditional cameras (CCD, CMOS) collect a large number of photons over a defined period (the exposure time) and convert the photon flux into an electrical charge through the photoelectric effect. The resulting voltage is then amplified and processed by an analog-to-digital converter, which produces the raw digital image output. 
This image capturing mechanism has several limitations, such as the full well capacity that constrains the number of photons each pixel can collect during the exposure time.
This results in saturated images when the incoming photon flux exceeds the full well capacity of the sensor.
Additionally, traditional sensors struggle in low-light scenarios, where a faint signal is hidden by the noise due to the analog amplification and the analog-to-digital converter, resulting in a low signal-to-noise ratio (SNR).
To reduce the noise at the pixel level, a common approach is to increase the exposure time, allowing the pixel to gather more photons, thus generating a stronger signal. Unfortunately, long exposure times introduce motion blur, as objects moving during the exposure period will appear blurred in the resulting image. 
In high dynamic range scenes, where some pixels receive low photon flux while others are brightly illuminated, the acquired image often contains very noisy pixels (due to low SNR) and saturated ones (because of the full well capacity being exceeded). Finally, because the photon flux is integrated over a predetermined exposure time, the frame rate of sequential acquisitions cannot be increased by post-processing.
%

%

Recent advancements in electronics have enabled the development of sensors sensitive to single photons, with extremely high readout speeds. Among these sensors, Single-Photon Avalanche Diodes (SPADs) \cite{spad_el, cusini2022historicalp1} can detect the arrival of a single photon by generating an exponential voltage gain from a single photoelectron. Time-Resolved SPADs (TR-SPADs) are SPAD sensors equipped with a fast time-to-digital converter, enabling precise timestamping of each photon’s arrival with picosecond-level accuracy. 
Because photon detections are naturally binary, SPADs are not affected by readout nor quantization noise \cite{morimoto20213}, resulting in a SNR which is only limited by the shot noise.
Thus, SPAD-based cameras are naturally suited for extremely low-light scenarios.
Additionally, the non-linear response of SPAD pixels enables sensing scenes with a very high dynamic range, without acquiring multiple exposures or additional hardware circuitry. In fact, theoretically, SPAD pixels only saturate with an infinite photon flux \cite{pf_spad, fossum2013modeling}. 
%
%
Moreover, by storing individual photon detections, the reconstruction frame rate can be adjusted in post-processing, offering novel opportunities to trade off motion blur and SNR.
Remarkably, TR-SPAD pixels can be read out asynchronously \cite{pf_spad, ip_spad}, that is, immediately after a photon is detected, bypassing the need to wait for the exposure time to end, leading to a virtually infinite frame rate.
%
%
These features make SPAD sensors revolutionary compared to traditional CCD-CMOS, sparking growing interest within the imaging community.

SPAD data, especially from time-resolved SPAD arrays, also pose unique challenges for computer vision and learning-based algorithms, paving the way for new research approaches.
Challenges include handling asynchronous photon arrivals, which prevents the direct use of conventional algorithms, as well as the growing data size. Even in mid-luminance scenes, raw data can easily contain thousands of timestamps per pixel, creating significant storage and processing demands. Existing reconstruction algorithms (e.g., \cite{pf_spad, ip_spad, yang2011bits}) address these issues by estimating the photon flux from the statistics of photon arrivals, resulting in a conventional digital image to be further processed by image processing algorithms. However, this approach obscures the asynchronous and online nature of the raw data and introduces an additional computational step, which may not be necessary for some vision tasks.

Although substantial research into models and methods for images captured by \emph{conventional} cameras has been conducted, the body of literature on SPAD image processing remains in its early stages of development, yet demonstrates significant potential for future developments \cite{sundar2022single, gupta2023eulerian, sundar2023sodacam, wei2023passive, jungerman2023panoramas, nousias2025opportunistic}.
One significant barrier to research on single-photon data processing is the lack of open-access datasets, which have been pivotal in the development of models for color images \cite{deng2009imagenet} and point clouds \cite{wu20153d}. 
This scarcity is a consequence of the fact that high-resolution asynchronous SPAD sensors are not yet commercially available (e.g., only 23 pixels in \cite{spad23}). 
For this reason, accurate SPAD simulators are of utmost importance to enable the development of new SPAD algorithms and hardware, as these provide a controlled environment for testing and developing algorithms without the limitations or costs associated with current physical equipment.

In this paper, we present a comprehensive approach for simulating SPAD acquisitions by modeling the physical noise source mechanisms involved in the the SPAD image formation pipeline.
Starting from a reference image, our simulator generates a photon detection stream as sensed by an asynchronous SPAD sensor, with adjustable photon rates to simulate various low-light conditions.
%
The simulator implements all the relevant SPAD readout modalities, namely asynchronous TR-SPAD, synchronous TR-SPAD (STR-SPAD) and the Quanta Image Sensor (SPAD-QIS).
We are the first to validate a simulator's accuracy on both asynchronous and frame-based SPAD data, through comparison with two commercial SPAD sensors from Pi Imaging \cite{spad23, spad512}. 
We analyze the distribution, rates, and counts of photon arrival with uniform flux, demonstrating that the simulator's output is realistic and coherent with real world SPAD data.

To promote further research in SPAD image processing, we release the SPAD-MNIST, a large dataset encompassing the three SPAD modalities, simulating the seminal MNIST dataset \cite{mnist_lenet} for image classification.
We believe that our dataset will spark interest in computer vision for SPAD sensors, since it represents the first large public SPAD dataset that can be used to train simple machine learning models for processing SPAD data.
In our experiments we primarily investigate two problems: \emph{i}) classification at very low light conditions and \emph{ii}) the effectiveness of pre-training a model on simulated data. We show that in very low light conditions, the noise in reconstructed fluxes hinder the image content in the reconstructed fluxes, hampering a simple task like digit recognition. In terms of synthetic data pre-training, we show that we can train a convolutional neural network (CNN) classifier exclusively on data from our simulator and we can successfully perform inference on SPAD images acquired by a commercial SPAD-QIS sensor, by establishing an ad-hoc setup for acquiring SPAD-QIS images at controlled lighting conditions. Even in these experiments the classification performance at varying illumination is assessed.


This paper extends our work on SPAD simulation \cite{suonsivu2025time} by introducing pixel-level non-uniformities. The extension includes modeling SPAD sensor characteristics, such as quantum efficiency, dark counts and afterpulsing, which are simulated individually for each pixel, to better align with real-world sensors. 
Additionally, we introduce the STR-SPAD and SPAD-QIS output modalities, which require a different quantization of the raw photon stream and reflect the most widely used SPAD-based sensor types.
We publicly release the SPAD-MNIST dataset including three modalities.
Finally, a novel validation analysis has been performed to assess the accuracy of our simulator.

\section{Passive Single-Photon Imaging}

Single-Photon Avalanche Diodes (SPADs) are photodetectors capable of detecting the incidence of individual photons on the pixel surface \cite{spad_el, cusini2022historicalp1, cusini2022historical}.
Similarly to traditional CMOS sensors, when a photon strikes a SPAD pixel, an electron is generated through the photoelectric effect. 
SPAD detectors can be arranged into large arrays to capture a wide field of view.
However, a SPAD pixel operates under reverse biasing above the breakdown voltage, resulting in an \emph{excess bias voltage}. 
This unique characteristic of SPADs makes a photoelectron initiate a self-sustaining avalanche of charges within the semiconductor, resulting in an exponential gain.
A specialized component, the quenching circuit, detects the avalanche and triggers a detection event. Since these events are digital pulses, SPAD sensors do not suffer from read-out nor quantization noise like CMOS sensors.
For these reasons, SPADs are highly effective in extreme low-light conditions, where CMOS experience severe performance degradation due to noise. 
%
Once the avalanche is detected, the quenching circuit resets the sensor to a stable state, preparing it for subsequent detections. The interval between avalanche detection and the sensor recovery is known as the \emph{dead time}, during which the sensor cannot register additional photons.

\subsection{SPAD modalities}

SPADs can be used either in an active or passive manner. 
In active SPAD-based sensing, SPADs are paired with an external light source~\cite{tontini2020numerical}, like a pulsed laser, which is utilized to perform time-correlated single-photon tasks.
SPADs have been used in fluorescence lifetime imaging microscopy (FLIM) \cite{async_spad_flim}, positron emission tomography \cite{lecoq2021sipm, jiang2019sensors}, Raman spectroscopy \cite{madonini2021single}, other biomedical applications \cite{wang2025fiber, bruschini2024review} and light detection and ranging (LiDAR) systems \cite{async_lidar, patanwala2021high, hu202132, lindner2018252}. 
This work focuses on passive SPAD-based imaging, where no active light source is used, but the SPAD passively measures incident light from the observed scene.
Passive SPAD sensors are attracting increasing interest for applications requiring extreme low-light sensitivity, very high dynamic range, and timely readout capabilities.

SPADs are classified mainly on the basis of their readout mechanism, which can be \emph{frame-based} or completely \emph{asynchronous}.
In the frame-based case, the readout is synchronous with an internal periodic clock signal, whose period is called \textit{integration time}.
Each frame stores the count of detected photons in a given pixel during an integration time. 
These counts are quantized with a certain bit depth, usually ranging from 1-bit to 12-bits. 
The most common data format for these sensors is 1-bit (binary) frames, often referred to as Quanta Image Sensor (QIS) \cite{fossum2016quanta}. 
In this format, a pixel is assigned a value of 1 if at least one photon is detected during the integration period; otherwise, it is set to 0.
To exploit the single-photon sensitivity and the extremely short integration time, 1-bit synchronous SPADs yield a sequence of binary frames, which can be regarded as a 3D volume of photon detections that occur during the measurement. Frame-based SPADs are capable of extremely high frame rates, exceeding 100,000 frames per second \cite{100kfps_spad}.

Moreover, SPAD sensors are often equipped with high-precision Time-to-Digital Converters (TDCs), which are used to timestamp the arrival of photons with a timing precision of pico- or nanoseconds and picoseconds timing jitter. 
In asynchronous SPAD sensors, each pixel operates independently from the others and is read out upon the occurrence of a detection event. 
These sensors are often referred to as asynchronous Time-Resolved SPADs (TR-SPADs).
Due to the asynchronous nature, each pixel generates a varying number of timestamps, resulting in raw data that cannot be directly organized into an array. Even uniformly illuminated pixels may generate different amounts of timestamps due to the stochastic nature of photon arrivals \cite{hasinoff2021photon}.
Although promising, asynchronous TR-SPADs are currently commercially available only in small arrays (e.g., 23 or 93 pixels in \cite{spad23, spad93}, or arrays of 32$\times$32 pixels in \cite{photonforce} and 64$\times$48 in \cite{novoviz}).

TR-SPAD arrays can also operate with synchronous readout, returning a frame that measures in each pixel the timestamp corresponding to the first photon detected during the exposure period. 
Synchronous TR-SPAD arrays are typically used in active imaging systems, where timestamps can be correlated with the timing of an illuminating source. 
For example, synchronous TR-SPAD arrays, combined with pulsed lasers, are core components of recent technologies such as LiDAR \cite{async_lidar} and FLIM systems \cite{spad_biphotonics}.

The choice of the SPAD modality is mainly application-dependent.
In low-light conditions, asynchronous TR-SPAD arrays are more attractive thanks to their higher data-efficiency than frame-based modalities, which would yield mostly zero-valued pixels.
This is because asynchronous readout enables detecting events to be triggered sparsely, whereas synchronous sensors are required to produce full frames even in the absence of photon detections.
Under bright illumination, it is often necessary to reduce the data rate of asynchronous TR-SPADs because of buffering and power constraints, especially when using large SPAD arrays.
In these cases, a synchronous array is preferred thanks to its implicit photon selection mechanism, which is a distinctive characteristic of 1-bit SPAD-QIS.
Finally, the timing information provided by the time-resolved sensors was shown to be useful in improving image reconstruction \cite{ip_spad, sync_tr_spad_rec}, but requires additional circuitry due to the per-pixel TDCs.

Regardless of the SPAD modality, specialized flux estimation algorithms \cite{yang2011bits, pf_spad, ip_spad, sync_tr_spad_rec} are required to reconstruct a digital image from SPAD raw data.
The reconstructed fluxes are then processed using traditional image processing and computer vision techniques, including denoising filters \cite{ip_spad} and visual recognition problems (e.g., classification by CNNs \cite{qis_classification}).
Directly processing raw photon streams for vision tasks is an exciting line of research that is  likely to attract growing interest.
In this paper, we describe a simulation framework for generating large datasets of synchronous SPAD-QIS and TR-SPAD, both asynchronous and synchronous, which is key for developing ad-hoc algorithms, including specialized learning-based models.

\subsection{Observation models}

Formally, we denote by $R \subseteq \mathbb{Z}^2$ the set of pixel coordinates in the SPAD array, and by $T \in \mathbb{R}_+$ the acquisition's exposure time.
%
The format of the raw data returned by each pixel $r \in R$ depends on the SPAD modality.
In binary synchronous SPAD-QIS having $M\in \mathbb{N}$ frames, each pixel  is read out periodically $M$ times producing a binary vector $B_r \in \{0,1\}^{M}$ of photon detections, where 1 means at least a photon was detected within the frame exposure time at pixel $r$.
Similarly, a synchronous TR-SPAD pixel records the arrival time of the first photon in each frame (relative to the start of the frame), thus it yields a vector $B_r \in [0,T]^M$ of photon time arrivals, with $B_r^f = T$ when no photon is detected in the $f$-th frame.
 %
In contrast, an asynchronous TR-SPAD exposed for a time interval of $T$, yields a stream of photon arrivals $\{x_r^i\}_{i=1}^{N_{r,T}}$,
where $N_{r,T} \in \mathbb{N}$ is the number of photons detected in pixel $r$ during the exposure $T$, and $x_r^i \in \mathbb{R}^+$ is the time of arrival of the $i$-th photons detected at the pixel $r$.
The statistical model of the measured quantities in all the three modalities is presented in Section~\ref{sec:simulator}.

\section{Related Work}

This section highlights the applications of simulated SPAD acquisitions and outlines the limits of existing low-light datasets.

\subsection{Datasets for single-photon imaging}

Arrays of time-resolved SPADs have only recently become available for purchase due to several manufacturing challenges.
Therefore, research on the processing of SPAD data has often been based on synthetic data, and only simplified experiments have been conducted in laboratory settings \cite{ip_spad, sync_tr_spad_rec}.

Simulators have been extensively used in the context of active, small-size, synchronous TR-SPADs arrays, with frames triggered by an illuminating laser.
In \cite{spad_planes,spikng_digits}, a simulator was used to generate acquisitions from a LiDAR based on TR-SPAD arrays, with the objective of developing models for feature extraction and classification. Another study \cite{3d_compressed, 3d_compressed_learned} utilized synthetic SPAD data to evaluate data compression techniques in LiDAR systems. 
In a similar vein, simulated detection timestamps \cite{rnn_fli} or arrival time histograms \cite{ann_fli} have been used to train neural networks for fluorescence lifetime imaging with SPAD sensors. 
These approaches focus on modeling the behavior of the illuminating source and its interaction with the SPAD sensors and require additional information, such as the 3D structure of the scene or the decay statistics of fluorophores.
In contrast, our simulator produces acquisitions of asynchronous photon detections for arbitrary-sized SPAD arrays, without any active illuminating source, and provide output as three different SPAD modalities.

Currently, there are no asynchronous SPAD arrays sufficiently large for time-resolved passive imaging; indeed, commercial sensors are very small (e.g., 23 pixels in \cite{spad23}).
Therefore, previous studies on passive asynchronous TR-SPAD \cite{pf_spad, ip_spad} utilized a TR-SPAD simulator to showcase the potential benefits of emerging TR-SPAD arrays over traditional CMOS cameras. 
These simulators process a photon flux as input, and simulate the image formation process of SPAD arrays.
Specifically, they simulate asynchronous streams of photon detections, modeling the physical response of each pixel and the noise corrupting acquisitions.
However, only a small number of simulated images were shared in the publications, and no large datasets have been made available to the community.
Moreover, the simulation procedures in these works have not been explicitly validated with real sensors. 
We have developed a simulator that converts any grayscale or color image into a photon flux calibrated according to a user-specified illuminance level.
Following established SPAD data formation models \cite{pf_spad, incoronato2021statistical, tontini2020numerical}, we generate photon detection streams for each pixel, starting from simulated fluxes, modeling pixel-level non-uniformities and noise sources of SPAD arrays.
We validate our simulator by comparing generated data to the response of two commercial SPAD sensors. 
To foster research for computer vision applications of SPADs, we simulate and release a large dataset for handwritten digit recognition. 
Additionally, we investigate the training of a CNN classifiers on synthetic data, with different illumination levels, and evaluate the trained classifier on real data.
We release the dataset for the three SPAD modalities: asynchronous time-resolved, synchronous time-resolved and single-bit SPAD-QIS.

Simulated datasets have been widely used in the QIS literature, including specific studies on SPAD-QIS.  
For instance, to address the image classification problem in low-light, large datasets of images have been simulated according to the QIS image formation model \cite{qis_classification}.
Simulations have been fundamental also for showcasing advanced reconstruction tasks, including color QIS imaging \cite{qis_color_rec} and online reconstruction \cite{qis_streaming_rec}.  
Directly simulating QIS data is way simpler than TR-SPAD ones, thanks to the lack of timing information and the standard frame-based data structure.
Our simulator is instead general, as it reproduces the statistics of photon detections in SPAD pixels, and then can easily generate synthetic SPAD-QIS acquisitions by quantizing raw TR-SPAD.

\subsection{Low-light datasets}

SPAD-based cameras have the potential to revolutionize low-light imaging applications, thanks to their single-photon sensitivity and absence of read-out noise.
In extremely low-light scenarios, traditional CMOS/CCD sensors yield severely degraded images due to their low photon sensitivity and strong influence of read-out noise.
Short exposure images are often enhanced by deep learning methods \cite{dl_low_survey_23}, which require large training sets of low- and high-light image pairs.
However, creating real-world datasets for training image enhancement models is a time-consuming and complex process, often requiring data at multiple exposures.
As a consequence, real low-light datasets may be small \cite{lol_dataset}, consist of video frames \cite{lliv_dataset}, or lack ground truth images \cite{exdark_dataset}.
Similar limitations also hinder the collection of large low-light datasets with SPAD cameras.

To overcome these difficulties, synthetic datasets are often used to train enhancement networks. 
The standard simulation method involves degrading a reference image through gamma correction and then adding noise, such as white Gaussian noise \cite{llnet}. 
However, these methods are based on simplifying assumptions that may not accurately reflect real-world conditions.
In contrast, we accurately simulate all noise sources involved in the acquisition procedure of a SPAD sensor.
Moreover, our simulator, which is specific for SPAD sensors, can generate acquisitions with arbitrary illumination levels and with user-specified SPAD characteristics.
Thus, our simulator allows us to generate realistic SPAD raw acquisitions, which can be used to produce arbitrary image datasets for low-light vision tasks, using appropriate image reconstruction algorithms.
In addition, simulated raw acquisitions can be used to directly investigate the design of novel processing methods for SPAD data in low-light scenarios.

\section{Simulator}
\label{sec:simulator}

In this section we illustrate our SPAD simulator and validate its performance with real single-photon detectors. 
The simulator is validated both against a real TR-SPAD sensor as well as 1-bit SPAD-QIS frames. We further use the simulator to generate the single-photon MNIST dataset with different SPAD modalities.

\subsection{Theoretical model for photon detection}
\label{ssec:theory}

The number of photon arrivals on an image sensor follows the Poisson statistics \cite{hasinoff2021photon}. Assuming an average (constant) photon flux $\phi$ hitting the sensor (photons per second hitting a pixel) for an exposure time $T$, the expected number of photon arrivals is $\phi T$. 
In particular, due to the nature of photon arrival, the probability of detecting $N$ photons during $T$ is determined by the Poisson process
\begin{equation}
    P(N) = \frac{(\phi T)^N \ \text{e}^{-\phi T}}{N!}.
\end{equation}
Under the Poisson assumption, the variance of the counts is equal to the expected counts. The photon arrival times are uniform randomly distributed, and the inter-photon times follow an exponential distribution with mean $1/\phi$.
This statistical model of photon arrivals describes the \emph{shot noise} and has been well validated in the literature \cite{timmermann1999multiscale, pf_spad, yang2011bits, wei2020physics}.

In addition to the natural variation in photon counts due to the Poisson process, the number of detected photons on the SPAD sensor is affected by several sensor-specific noise factors, whose parameters are listed in Table~\ref{tab:symbols}.
First, as in CMOS/CCD photodiodes, the quantum efficiency of each pixel determines the probability of a photon to be absorbed in the semiconductor medium. With SPAD operating in the Geiger mode, the quantum efficiency is 
accompanied by the avalanche triggering probability, i.e. the probability that a photon absorbed in the active region of the SPAD creates an avalanche.
Considering these two factors together, it is common to describe the probability that an impinging photon is detected by the \emph{photon detection probability} (PDP). 
The PDP depends on the wavelength of the incoming light flux with a peak probability in the visible wavelength range for silicon SPADs \cite{bronzi2015spad}.
Second, the \emph{fill factor} (FF) is an important parameter of SPADs 
that describes the ratio of the active (photo-sensitive) area to the pixel total area. 
Combining all these factors, the overall sensitivity of SPAD arrays is determined by the \emph{photon detection efficiency} (PDE), defined as $PDE = FF \cdot PDP$.
Commonly, microlenses are used for enhancing the PDE \cite{spad_biphotonics}, increasing the total photon detection efficiency or sensitivity of SPAD arrays. 

After a charge carrier avalanche is created by an impinging photon, the SPAD pixel undergoes a reset process, causing the SPAD to become temporarily inactive for a time duration $\tau_d$, referred to as the \emph{dead time}. During the dead time, the carrier avalanche is quenched and the pixel voltage is charged back to the operating (excess bias/over) voltage. The duration of the dead time is typically in the low nanosecond scale, up to 100s of nanoseconds \cite{bronzi2015spad, gramuglia2021low}.

During an avalanche, some charge-carriers may be trapped by defects in the semiconductor and later released after the photon detection, possibly triggering a secondary avalanche known as an \emph{afterpulse} \cite{morimoto2021scaling}. The probability of an after-pulse, $P_{ap}$, depends on the pixel design and is typically in the order of $\sim$1\%~\cite{bronzi2015spad}. 
The time delay from the first photon to the afterpulse is often characterized by a single exponential distribution with time constants $\bar{\tau}$ ranging from a few up to 1000s of nanoseconds \cite{giudice2003process}. Depending on material and pixel design, the total response may however consist of several trapping levels with different lifetimes and probabilities \cite{da2011real}.

SPADs may exhibit counts in complete darkness as well. These spurious counts, referred as \emph{dark counts}, are uncorrelated in time and follow the Poisson distribution, and  they result typically from thermal excitation. The Dark Count Rate (DCR) is a critical factor for low-noise SPAD development and the DCR typically spans from 10 to 1000s of Hz for silicon SPADs, depending on various sensor specific and external factors such as the excess bias voltage and temperature \cite{xu2017comprehensive, sicre2021dark,morimoto2021scaling, bronzi2015spad}.

Lastly, 
the detected photon arrival time is affected by some stochastic factors. Depending on avalanche triggering process, the photon timestamps typically exhibit \emph{jitter} $\Delta \tau$ around an expected delay. 
Assuming that the photon absorption occurs prominently in the depletion region of the pixel, where the electric field is the highest, the jitter can be modeled by a normal distribution \cite{sun2019simple}. However, the avalanche can also be triggered by a photon absorption in the neutral region of the SPAD by a carrier diffusion into the depletion region. This process is slower and less likely to occur, adding an exponential tail to the jitter distribution \cite{sun2019simple}. The jitter can play a crucial in active imaging modalities such as FLIM and time-of-flight, but, it has no substantial effect on the photon arrivals for the passive imaging.
The photon arrival times, affected by the jitter, can then be recorded by a TDC whose temporal resolution $\Delta t$ is determined by the clock signal.

\begin{table}[t]
    \centering
    \caption{SPAD parameters, symbols and reference values (see Refs.~\cite{cusini2022historical, bronzi2015spad, spad_biphotonics, charbon20183d}).}
    \vspace{2mm}
    {\small
    \begin{tabular}{l c c}
        Parameter & Symbol & Typical values \\
        \hline
        Exposure time & $T$ & ns--s \\
        Pixel area & $A_p$ & 25--400 \textmu m$^{2}$ \\
        Excess bias voltage & $V_{ex}$ & 1--10 V \\
        Photon detection prob. & $q$ & 30--70 \% (peak) \\
        Fill factor & FF & $<$10--100 \% \\
        Dead time & $\tau_d$ & 1--100 ns \\
        Afterpulsing probability & $P_{ap}$ & $<$1 \% \\
        Dark count rate & $\phi_d$ & 1--1000 Hz \\
        Timing jitter & $\Delta \tau$ & 20--500 ps \\
        Timing resolution & $\Delta t$ & 5--100 ps \\
        \hline
    \end{tabular}
    }
    \label{tab:symbols}
\end{table}

\subsection{Simulation pipeline}
\label{sec:scene_to_photons}

\begin{figure}[t]
    \centering
    \includegraphics[width=\columnwidth]{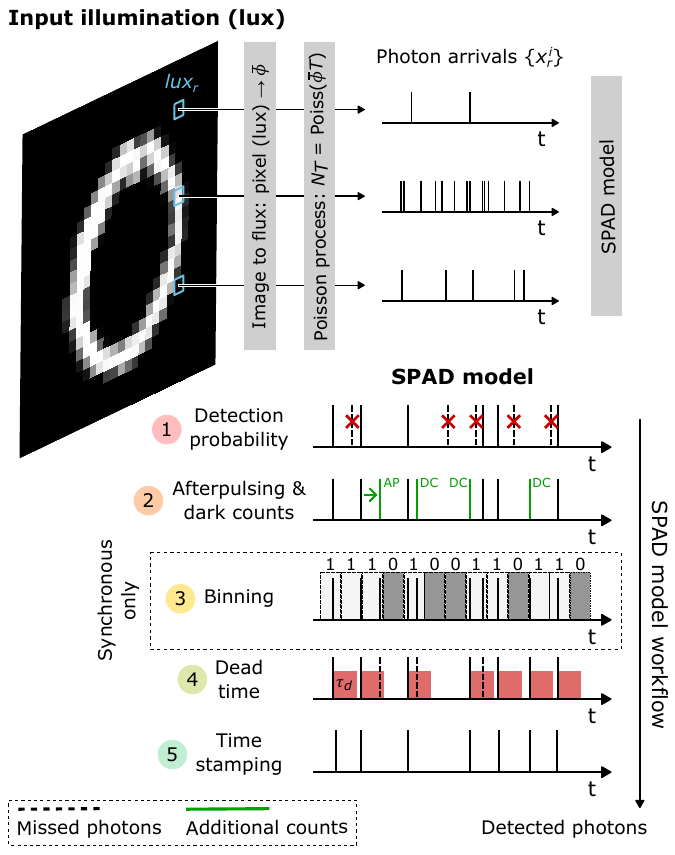}
    \caption{Illustration of the SPAD simulation model. The SPAD model is applied pixel-wise in the simulations. Notice that the QIS frame exposure time is typically much longer than the dead time of the SPAD (i.e. not plotted in scale).}
    \label{fig:schematic}
\end{figure}

The SPAD simulation process is illustrated in Fig.~\ref{fig:schematic} and described in detail in Alg.~\ref{alg:pipeline}, where a grayscale image is used as the simulator input. The first block of the simulator involves turning each pixel value into a photon flux hitting each pixel, followed by a perturbation due to the Poisson process in the second block. The resulting photon streams $\{ x_r^i\}$ are subsequently influenced by SPAD-specific noise factors described in Section~\ref{ssec:theory}.
In the set of arrivals $\{ x_r^i\}$, $x_r^i$ denotes the arrival time of $i$-th photon on pixel coordinate $r \in R$. The simulation process described below is applied individually to each pixel $r$ with one-to-one mapping from the input image,
without considering spatial correlations such as optical crosstalk \cite{rech2008optical}, which are not prominent in the commercial SPAD imagers.
Then, we can simulate SPAD data in a per-pixel manner, where we omit the pixel index $r$ in the following for the notation sake.
Thus, the simulation can be performed in parallel for each pixel.

To turn the input image into a photon flux, we introduce an effective reference lux level $R_{lux}$ on the sensor level such that the range of the input image (e.g., [0, 255] for an 8-bit image) is scaled to a range [0, $R_{lux}$]. This scaling is performed by the ``Image to flux" block in Fig.~\ref{fig:schematic} and the scaled image is denoted by $I_{lux}$.
We consider a monochromatic illumination, and the image lux values $I_{lux}$ are translated to power values per pixel $I_W$ by
\begin{equation}
    I_W = \frac{A_p^{act} I_{lux}}{K},
    \label{eq:IW}
\end{equation}
where $K$ = 683~lm/W is luminous efficacy at 555~nm \cite{arecchi2007field}, and $A_{p}^{act}$ is the \emph{active} pixel area expressed in m$^2$. Note that the active pixel area is determined by $A_{p}^{act} = FF \cdot A_{p}$, where $FF$ is the array fill factor and $A_{p}$ is the total area of the SPAD pixel.
The expected photon flux on the pixel is calculated as 
\begin{equation}
    \bar{\phi} = \frac{I_W}{E_{ph}},
    \label{eq:phibar}
\end{equation}
where $E_{ph}(\lambda) = hc/\lambda$ is the energy of a single photon at wavelength $\lambda$, and $h$ and $c$ are the Planck constant and speed of light, respectively. 

The second block of the simulator, referred to as ``Poisson process", defines
the actual photon number $N_T$ reaching each SPAD pixel during a time interval $T$. $N_T$ is determined by drawing a realization from the Poisson process: \mbox{$N_T \sim \mathrm{Poiss}(\bar{\phi} \cdot T)$}, representing the shot noise. The photon arrival times are drawn from a uniform distribution over the time interval $T$, generating the stream $\{ x^i\}$ of photon arrival times (lines 4--5 in Alg.~\ref{alg:pipeline}). 

Next, the incident photon stream $\{x^i\}$ enters the ``SPAD model" block where photons are removed from the stream with a probability $(1-PDP)$ (Step 1 in Fig.~\ref{fig:schematic}, lines 6--9 in Alg.~\ref{alg:pipeline}).
Additional, spurious photon arrivals due to dark counts are added to the photon stream in each pixel following the Poisson process (lines 10--12).
In Alg. \ref{alg:pipeline} and Table \ref{tab:symbols} we denote the DCR and PDP as $\phi_d$ and $q$, respectively.
Afterpulsing effects are introduced by adding, with probability $P_{ap}$, a second detection shortly (within nanoseconds) after a detected photon or a dark count (Step 2, lines 14--18). The time delay of the afterpulse (line 17) with respect to the primary detection is drawn from an exponential distribution with a de-trapping lifetime $\bar{\tau}$, shifted by the SPAD pixel dead time $\tau_d$.
The stream of recorded photons $\{x^i\}$ is fed as input to the following steps to generate the three different SPAD modalities, as described in Table~\ref{tab:mods}.

For the both synchronous SPAD modalities (QIS-SPAD, STR-SPAD), the simulation process continues by binning over frame exposure time $T_f$ (Step 3, lines 20-21). For the QIS-SPAD, if one or more photons are detected during $T_f$, the pixel value is set to 1; otherwise to 0.
Formally, the binning operation at line 21 for the QIS-SPAD data and frame $f$ is expressed as \mbox{$B_f = \text{min}(1, \ \text{len}(\{x^i\}))$}, for \mbox{$x^i \in [f\cdot T_f, (f+1)\cdot T_f)$.}
Operator len() refers to the number of photons in the stream.
The \emph{synchronous} time-resolved (STR-SPAD) simulations include the binning but the $B_f$ contains the timestamp (Step 5) of the first photon arrival within the frame exposure.

For the fully \emph{asynchronous} TR-SPAD outputs, the binning is omitted and the next step includes the application of the dead time 
by removing the photon arrivals that occur with a time delay shorter than the dead time $\tau_d$ when compared to last recorded photon (solid lines in Step 5, $y^{end}$ on lines 24--29).
Finally, the timestamps (Step 5, lines 30--32) are saved with a full or limited temporal resolution provided by the TDCs.

\begin{table*}[t]
    \centering
    \caption{Simulator steps included to different SPAD modalities.}
    \vspace{2mm}
    {\small
    \begin{tabular}{x{2cm}|x{2.2cm}|x{2.2cm}|x{2.2cm}|x{2.2cm}|x{2.2cm}}
        Modality & PDP & AP \& DC & Binning & Dead time &  Time-stamping \\
        \hline
        TR-SPAD & {\color{green}\ding{51}} & {\color{green} \ding{51}} &{\color{red} \ding{55}} & {\color{green}\ding{51}} & {\color{green}\ding{51}}\\
        STR-SPAD & {\color{green}\ding{51}} & {\color{green}\ding{51}} & {\color{green}\ding{51}} & {\color{red}\ding{55}} & {\color{green}\ding{51}} \\
        QIS-SPAD & {\color{green}\ding{51}} & {\color{green}\ding{51}} & {\color{green}\ding{51}} & {\color{red}\ding{55}} & {\color{red}\ding{55}} \\
    \end{tabular}
    }
    \label{tab:mods}
\end{table*}

\begin{algorithm}[!ht]
\caption{Simulator pipeline schematic}\label{alg:pipeline}
\begin{algorithmic}[1]
\INPUT $I_{lux} \in [0, R_{lux}]$ \COMMENT{Scaled input image}
\INPUT $T > 0$ \COMMENT{Exposure time}
\INPUT $q \in [0,1]$ \COMMENT{Photon detection probability}
\INPUT $\phi_d \geq 0$ \COMMENT{Dark count rate}
\INPUT $\tau_d > 0$ \COMMENT{Dead time}
\INPUT $P_{ap} \in [0,1]$ \COMMENT{Afterpulsing probability}
\INPUT $\bar{\tau} > 0$ \COMMENT{Trap lifetime}
\INPUT $\Delta t \geq 0$ \COMMENT{Temporal resolution}
\INPUT $N_p > 0$ \COMMENT{Number of pixels}
\INPUT Modality \COMMENT{TR-SPAD/STR-SPAD/QIS-SPAD \hspace*{-1em}}
\STATE Calculate luminous power $I_W$ \COMMENT{Eq. \eqref{eq:IW} \hspace*{-1em}}
\STATE Calculate photon flux $\bar{\phi}$ \COMMENT{Eq. \eqref{eq:phibar} \hspace*{-1em}}
\FOR[Omit pixel index $r$ \hspace*{-1em}]{pixel index $r=0$ to $N_p$}
    \STATE $N_T \gets \text{Poiss}(\bar{\phi} \cdot T)$ \COMMENT{Number of photons \hspace*{-1em}}
    \STATE $\{x^i\} \gets \{U(0, T) \}^{1 \times N_T}$ \COMMENT{Photon arrivals \hspace*{-1em}}
    \FOR{photon arrival $j=1$ to $N_T$}
        \STATE $u \gets U(0, 1)$ \COMMENT{Temporary random variable \hspace*{-1em}}
        \IF{$u < 1-q$}
            \STATE $\{x^i\} \gets \{x^i\} \setminus x^j$ \COMMENT{Not detected}
        \ENDIF
    \ENDFOR
    \STATE $N_{DC} \gets \text{Poiss}(\phi_d \cdot T)$ \COMMENT{Number of dark counts \hspace*{-1em}}
    \STATE $\{x_d^i\} \gets \{U(0, T) \}^{1 \times N_{DC}}$ \COMMENT{Dark counts \hspace*{-1em}}
    \STATE $\{x^i\} \gets \{x^i\} \cup \{x_d^i\}$ \COMMENT{Add dark counts \hspace*{-1em}}
    \STATE $N \gets \text{len}(\{x^i\})$ \COMMENT{Compute number of detections \hspace*{-1em}}
    \FOR{detection event $j=1$ to $N$}
        \STATE $u \gets U(0, 1)$ \COMMENT{Temporary random variable \hspace*{-1em}}
        \IF{$u < P_{ap}$}
            \STATE $\tau^j \gets \text{Exp}(1/\bar{\tau}) + \tau_d$ \COMMENT{Delay for afterpulse \hspace*{-1em}}
            \STATE $\{x^i\} \gets \{x^i\} \cup x^j + \tau^j$
        \ENDIF
    \ENDFOR
    \STATE $\{x^i\} \gets \text{Sort}(\{x^i\})$ \COMMENT{Sort detection events \hspace*{-1em}}
    \IF[Synchronous mode (STR/QIS) \hspace*{-1em}]{frame-based}
        \STATE $B \gets \text{Binning}(\{x^i\}, T_f, \text{modality})$
        \RETURN $B$ \COMMENT{Return frames}
    \ELSE[Asynchronous mode (TR) \hspace*{-1em}]
        \STATE $N \gets \text{len}(\{x^i\})$
        \STATE $\{y^i\} \gets x^1$ \COMMENT{Initialize new stream \hspace*{-1em}}
        \FOR{detection event $j=2$ to $N$}
            \IF[Record detection \hspace*{-1em}]{$x^{j} - y^{end} > \tau_d$} 
                \STATE $\{y^i\} \gets \{y^i\} \cup x^j$
                \STATE $y^{end} \gets x^j$ \COMMENT{Update last detected}
            \ENDIF
        \ENDFOR
        \IF{limited temporal resolution}
            \STATE $\{y^i\} \gets \text{Round}(\{y^i\}, \Delta t)$
        \ENDIF
        \RETURN $\{y^i\}$ \COMMENT{Return photon stream}
    \ENDIF
\ENDFOR
\end{algorithmic}
\end{algorithm}


\subsection{Simulator validation}

\begin{figure}[!b]
    \centering
    \includegraphics[width=\columnwidth]{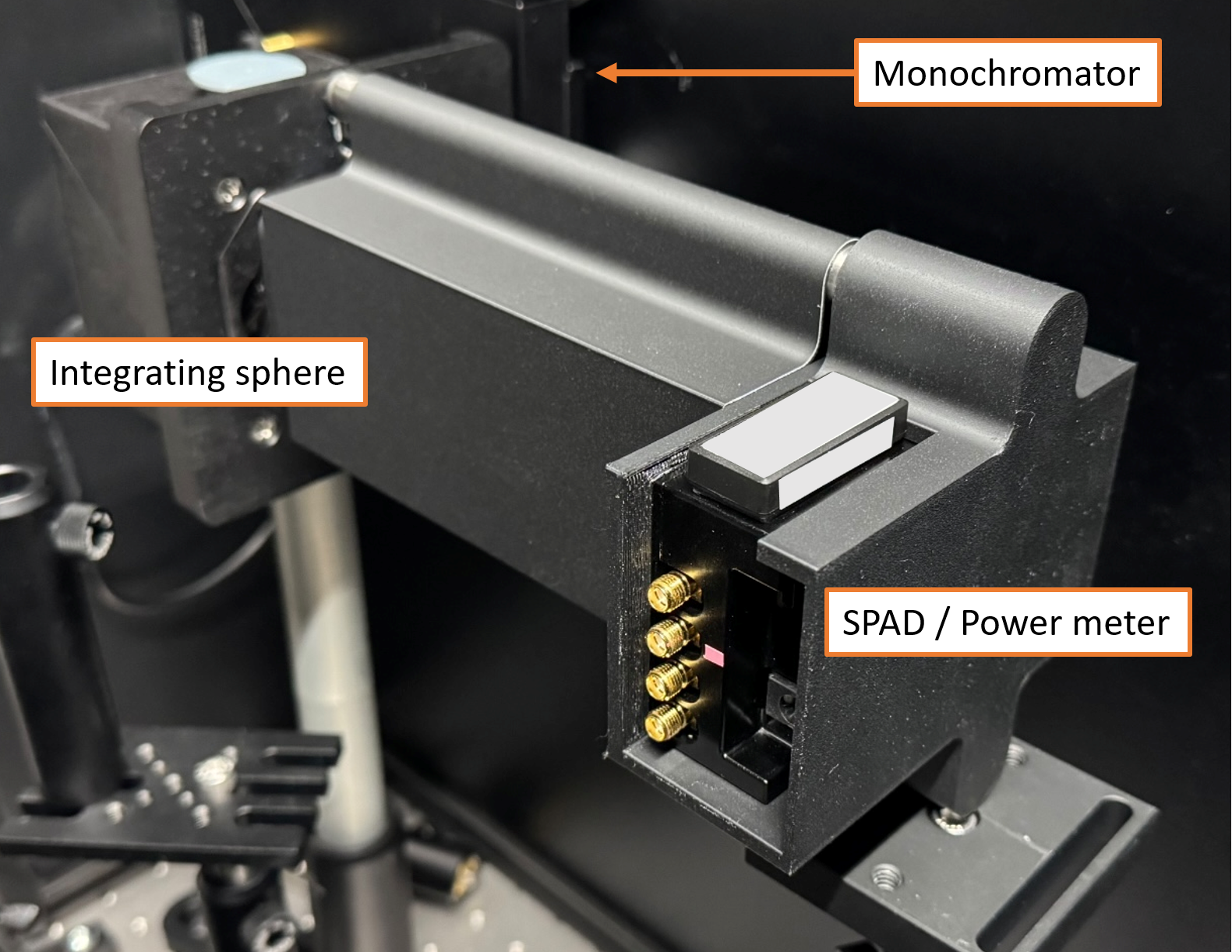}
    \caption{Experiment setup consisting of a tunable narrowband light source, integrating sphere and the SPAD sensor.}
    \label{fig:monochromator}
\end{figure}

In order to validate the accuracy of our simulator, we conduct a comprehensive series of experiments, in which we compare the outputs of our simulator to data acquired from commercial SPAD sensors. Our analysis encompasses both asynchronous and frame-based single-photon data acquired at different wavelengths. We use a tunable narrowband light source (Labsphere QES-1000 combining a plasma lamp and an incandescent source, followed by a monochromator, a neutral density filter wheel and an integrating sphere) 
to illuminate the sensor and a reference optical power meter (Thorlabs S130C) with a uniform flux. The reference power meter is used for estimating the incoming photon flux hitting the SPAD sensors.
Measurements were acquired with diffused light at $\sim$20~cm from the integrating sphere where the SPAD sensor and power meter were positioned. We set the illumination to two wavelengths: 500 and 700~nm. Notice that the intensity of the light source is $\sim$5 times higher at 500~nm compared to 700~nm. The experimental setup is shown in Fig.~\ref{fig:monochromator}.

\begin{figure}[!t]
    \centering
    \includegraphics[width=\columnwidth]{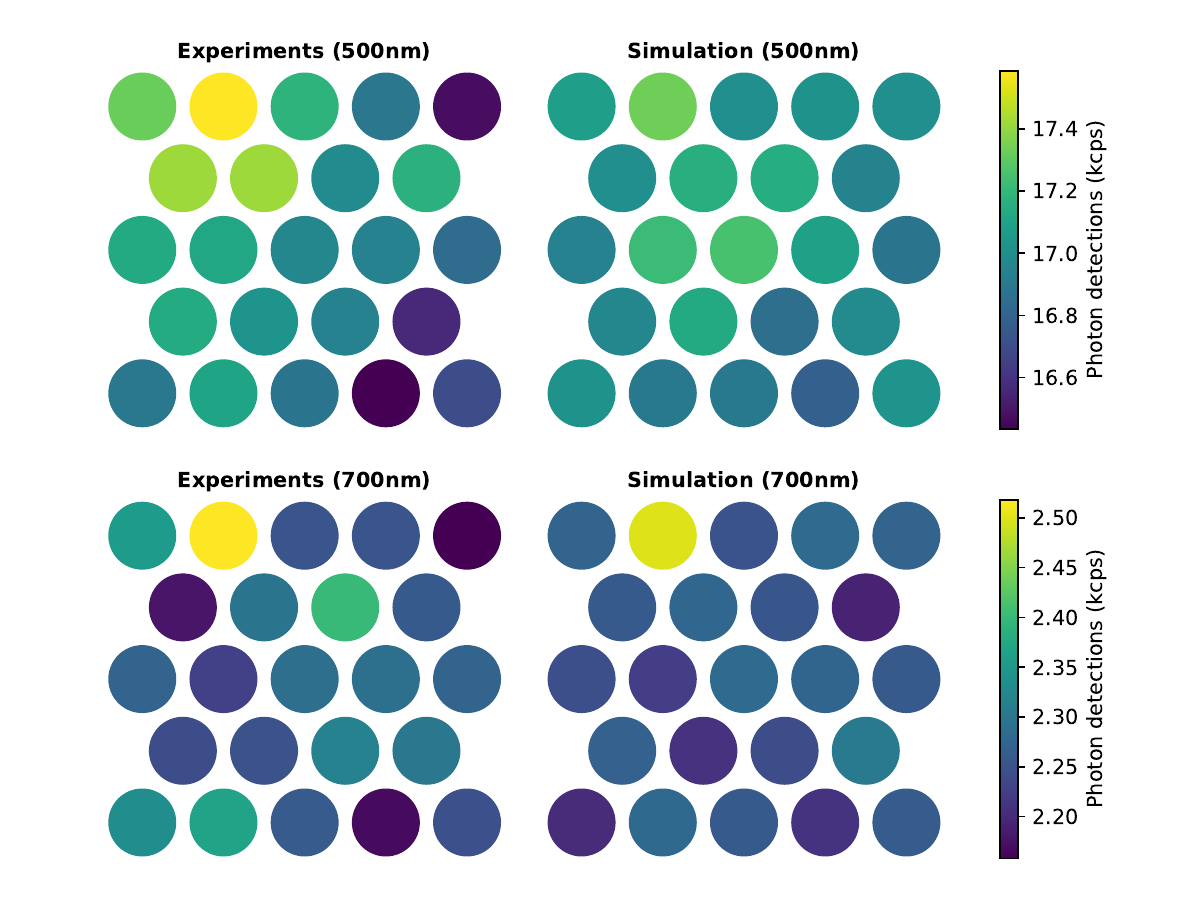}
    \caption{SPAD23 simulator and experiments comparison at intermediate flux. The mean photon flux is 17.0 kcps in both experiments and simulations for 500 nm, and 2.28 kcps and 2.27 kcps for 700 nm, respectively.}
    \label{fig:spad23val}
\end{figure}

We validate our simulator against a TR-SPAD sensor: the SPAD23 from Pi Imaging \cite{spad23}, containing 23 time-resolved pixels. The SPAD pixels are arranged into a hexagonal pattern, as shown in Fig.~\ref{fig:spad23val}. The SPAD is operated at 5~V excess bias and we collect 10 measurements of 100~ms duration each.
We configured our simulator at an estimated PDE of 42.2\% at 500~nm and 16.2\% at 700~nm for the pixel area of 458~\textmu m$^2$ provided by the Pi Imaging datasheet.
We include a pixel-wise DCR into the simulator based on the measured dark counts. The simulations were performed with a spatially uniform PDE and by setting the dead time to 50~ns 
and the afterpulsing probability 0.1\% as reported in the Pi Imaging datasheet. The de-trapping lifetime was set to 25~ns.
Notice that the SPAD23 sensor is equipped with microlenses that increase the effective fill factor from native 23.5\% to near 100\%.

Fig.~\ref{fig:spad23val} shows the results for the simulator validation against the SPAD23, illustrating the detected photon rate (counts per second, cps) at the two wavelengths (500~nm on the top row; 700~nm on the bottom row).
The measured input illumination on the SPAD was approximately 3 and 0.8 nW/cm$^2$ for 500~nm and 700~nm, respectively.
For both wavelengths, the simulations match the experimental results, with less than 1\% difference in the mean photon counts.
The simulations assume a uniform PDE across the SPAD array, though real sensors typically exhibit pixel-to-pixel variations, leading to larger variability in photon counts.

We also validate frame-based images from our simulator against a QIS-SPAD from Pi Imaging \cite{spad512}, referred to as SPAD512$^{2}$. This sensor outputs the photon detections in frames with desired amount of bits representing the photon counts. Supported bit-depths for this sensor range from 1 to 12 bits. 
In our validation, we focus on the 1-bit frames, which is perhaps the most interesting data format, as at high frame rate this preserves as much of the original light signal as possible.
By exposing the sensor to a uniform (spatial and temporal) flux, we can collect single-photon data and analyze the distribution of the detected photons, without having undesired non-uniformities such as flickering in the light stream.

We operate the SPAD512$^{2}$ with an excess bias of 6~V and integration time of 1 or 10~\textmu s per frame, and collect $M$=10,000 binary frames. The sensor is capable of capturing binary frames with up to 100,000 frames per second.
In the simulations, we assume a spatially uniform PDP. 
Considering a measured incoming flux from the light source, we estimate the PDP of the sensor to be 44.4\% at 500~nm and 20.7\% at 700~nm for an active pixel area 36~\textmu m$^2$. The DCR in the simulations was estimated by measuring the dark counts of the sensor acquiring binary raw data with a closed shutter. The native fill factor of the sensor is $\sim$13\% and it does not have microlenses.

\begin{figure}[!t]
    \centering
    \includegraphics[width=\columnwidth]{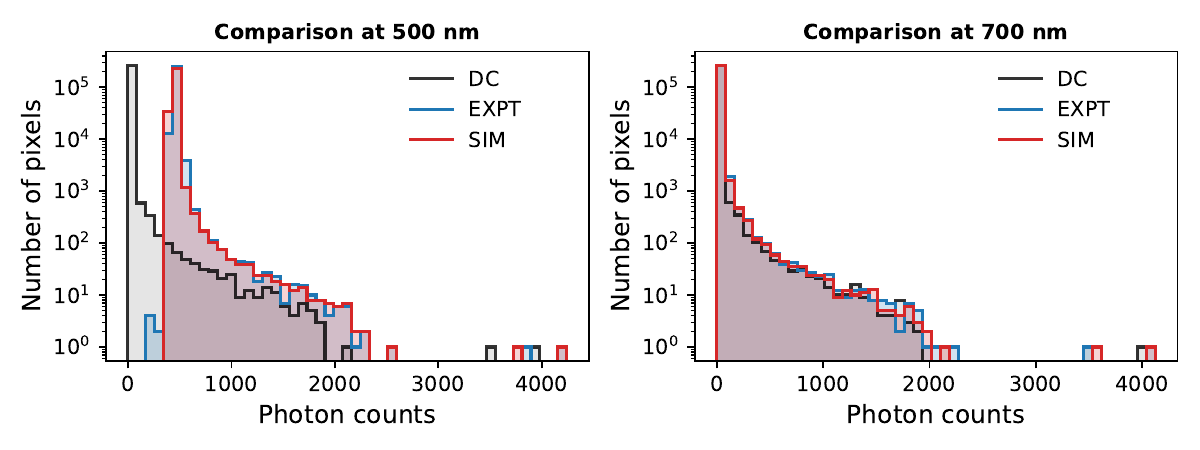}
    \caption{SPAD512$^2$ simulator and experiments count histogram comparison at low photon counts. 
    At 500~nm, the mean photon counts are 470 and 457 for the experiments (EXPT) and simulations (SIM), respectively, and for 700~nm the mean photon counts are 62 and 60.
    The counts are over 10000 binary frames with 1~\textmu s integration time. The histograms also show the measured dark counts (DC) with a mean photon count of 2.6, with 265 (20) counts per second mean (median) DCR.}
    \label{fig:spad512val}
\end{figure}

\begin{figure}[!ht]
    \centering
    \includegraphics[width=\columnwidth]{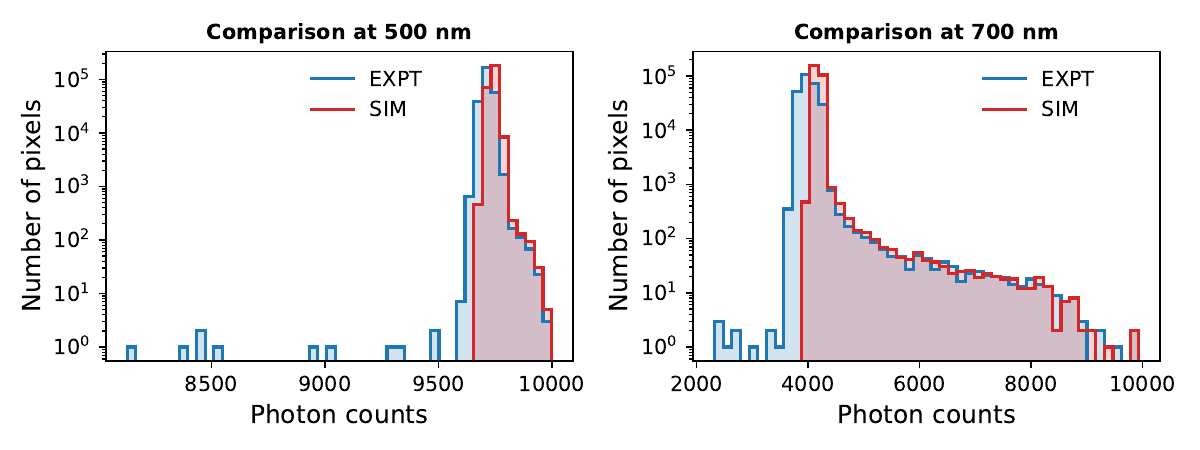}
    \caption{SPAD512$^2$ simulator and experiments count histogram comparison at high photon counts.
    At 500~nm, the mean photon counts are 9715 and 9740 for the experiments (EXPT) and simulations (SIM), respectively, and for 700~nm the mean photon counts are 4011 and 4184.
    The counts are over 10000 binary frames with 10 \textmu s integration time. Note that compared to Fig. \ref{fig:spad512val}, the overall incoming flux is also higher. The simulations currently consider all pixels to have a uniform PDP, and the effect of "lazy pixels" can be seen in the experiments.}
    \label{fig:spad512val2}
\end{figure}

Fig.~\ref{fig:spad512val} shows the photon count comparison between the simulations and data acquisition from SPAD512$^{2}$ at low light with 1~\textmu s integration time per binary frame.
The measured input illumination on the SPAD was approximately 120 and 20 nW/cm$^2$ for 500~nm and 700~nm, respectively.
We also plot the measured dark counts (DC) in the absence of illumination. At 500~nm illumination, the total counts measured and simulated are significantly higher compared to the dark counts only. While the overall correspondence between the simulations and experiments is excellent, a small number of pixels in the experiments detect 
a low number of photons compared to the majority of the pixels (left edge of the histogram).
This phenomenon
is apparent in Fig.~\ref{fig:spad512val2} as well, and is due to pixels having a reduced PDP, referred to as \emph{dead} or \emph{lazy pixels}. The explanation for this behavior is likely related to a variation in the excess bias voltage of those (group of) pixels \cite{antolovic2015nonuniformity, xu2017design}. 
At 700~nm illumination, the sensitivity (PDP) of the SPAD512$^{2}$ is reduced compared to 500~nm, and furthermore the intensity of the source is lower as well at this wavelength. Thus, the main contributor to the detected counts in these measurements and simulations is the dark counts.
At both wavelengths, we can see a few pixels with very high photon counts. These pixels are called \emph{hot pixels} (or screamers) as their DCR is high. 
The SPAD23 also contains a very apparent hot pixel, as shown
in Fig.~\ref{fig:spad23val} with a significantly higher photon count.

Next, we increased the brightness of the source by a factor of $\sim$10 and repeated the experiments on the SPAD512$^{2}$. We further increased the integration time per frame up to 10~\textmu s. Thus, Fig.~\ref{fig:spad512val2} shows the photon count comparison between the simulations and experiments for significantly higher photon counts compared to Fig.~\ref{fig:spad512val}.
From these photon count histograms, the presence of lazy pixels is more evident. 
We also observe a slight shift among the two distributions that may be due to uncertainty in the power estimation for the simulations, and a broader peak in the experiments due to pixel-wise PDP variation.

In conclusion, we validated the simulator for both asynchronous and synchronous modalities. Our experiments show that the simulator can accurately reproduce data acquired from the two sensors within the uncertainties of the experimental setup. 
The simulation accuracy could be further improved by the application of a pixel-wise photon detection probability, as we now consider a uniform PDP across both modeled sensors.

\section{Dataset}

\begin{figure}[!b]
    \centering
    \includegraphics[width=\columnwidth]{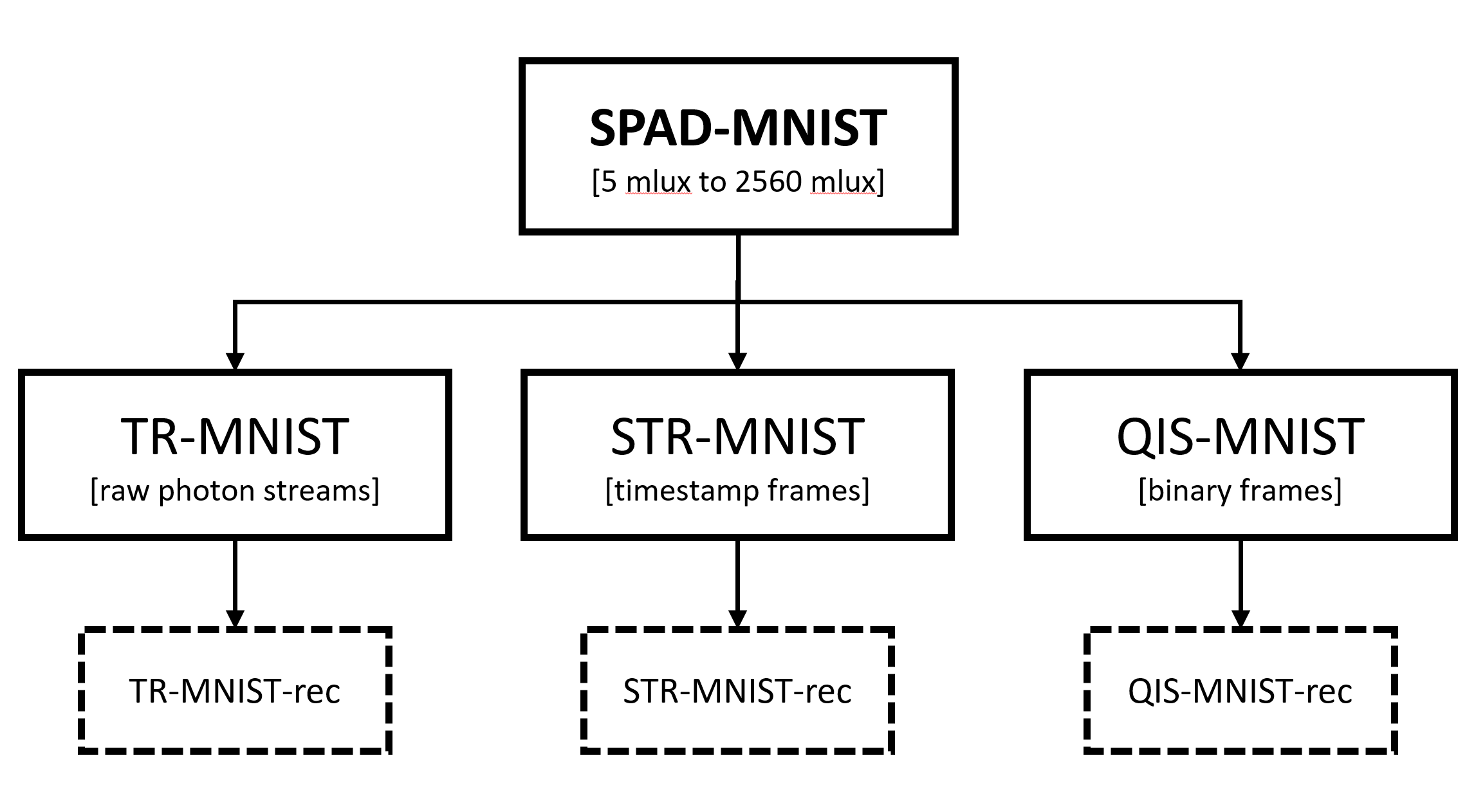}
    \caption{Illustration of the SPAD-MNIST dataset and its subsets.}
    \label{fig:spadmnist}
\end{figure}

To address the lack of publicly available SPAD data, in this paper we present a single-photon MNIST dataset which we call SPAD-MNIST. The dataset consists of three sub-datasets, which are described in detail in Section \ref{ssec:datadesc}. The structure of the dataset is illustrated in Fig.~\ref{fig:spadmnist}. Flux reconstruction algorithms to get from raw photon data to images are described in Section \ref{ssec:recon}.
The full dataset and sample scripts for processing the data can be downloaded from the project page {\small \url{https://boracchi.faculty.polimi.it/Projects/SPAD-MNIST.html}}.

\subsection{Dataset description}

\label{ssec:datadesc}

The SPAD-MNIST dataset has been generated from the original MNIST \cite{mnist_lenet} dataset, by simulating raw photon streams with our simulator and processing the streams according to the different SPAD modalities as described in Section~\ref{sec:scene_to_photons}.
We selected MNIST as the base for our dataset, as it has played a crucial role in computer vision, especially for image recognition. It is a simple image classification dataset, consisting of labeled 28$\times$28 grayscale images where each image contains a hand-written digit (from 0 to 9). The simplicity of MNIST, due to its small image size and limited number of classes, stimulated interest in learning-based computer vision research, even with the limited computational resources available at the time.
We generate the SPAD-MNIST dataset using multiple extreme low-light illumination levels, ranging from 5 mlux to 2560 mlux.

As illustrated in Section~\ref{sec:simulator}, our simulator enables the generation of realistic single-photon data.
In the SPAD-MNIST simulations, we consider a SPAD array of size 28$\times$28, under a varying input photon flux, with an active area of 25~\textmu m$^2$ and 100\% fill factor, photon detection probability of 50\%, and DCR of 100~Hz. The dead time is set to 50~ns, afterpulsing probability to 0.5\%, and 
and de-trapping lifetime to 25~ns.
The sensor is exposed for a total duration of 1~ms in all simulations. 
The simulator was implemented using Matlab (and executed on an Intel i9-9900 CPU, 64 GB RAM) and the run time was around 40 minutes for one lux level (70,000 samples), including loading the MNIST images, saving the TR-SPAD data and reconstructions.

We provide data for the asynchronous SPAD modality in both raw photon streams (TR-MNIST) as well as the reconstructed photon flux images (TR-MNIST-rec). Section~\ref{ssec:recon} describes the used photon flux reconstruction methods.

Compared to \cite{suonsivu2025time}, we introduce the QIS-MNIST that includes reconstructed binary frames. These frames are generated from the original photon stream simulations (1~ms duration) and quantized into 100 binary frames with 10 \textmu s integration time per frame for all illumination levels.
We also introduce a dataset for synchronous TR-SPAD modality (STR-MNIST) where instead of the binary values, the frames consist of the timestamps of first photon arrivals within a frame, relative to the start of the frame's exposure.

The SPAD-MNIST dataset represents a benchmark for future deep learning research for processing low-light photon streams, as well as, for designing novel flux estimation algorithms and image classification tasks using SPAD cameras.

\subsection{From photon streams to reconstructed images}
\label{ssec:recon}

\begin{figure}[!t]
    \centering
    \includegraphics[width=\columnwidth]{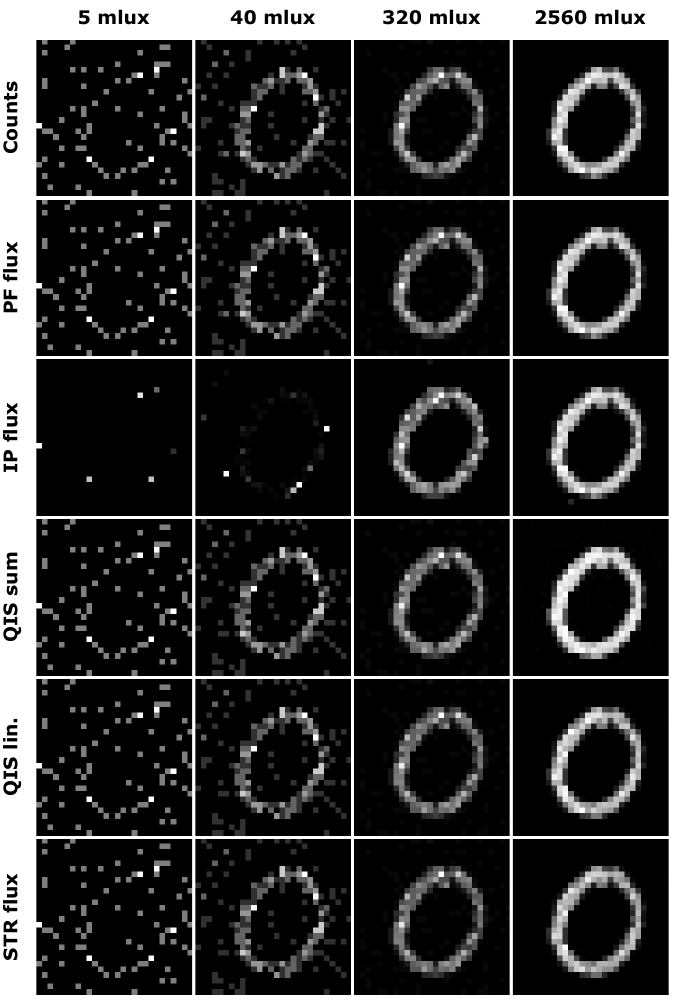}
    \caption{Samples of flux reconstructions at different lux levels. The maximum of the normalized fluxes (see text) has been clipped to a value of 20 for visualization.}
    \label{fig:samples}
\end{figure}

In order to visualize the raw photon data in image format, the photon flux needs to be estimated to form an image.
Examples of flux reconstructions are shown in Fig.~\ref{fig:samples}.
The TR-MNIST-rec dataset contains photon flux reconstruction images obtained from the synthetic raw photon data. It includes the following three flux reconstruction methods \cite{pf_spad, ip_spad}:

\begin{subequations}
\begin{equation}
\widehat{\Phi}_{C} = \frac{N_T}{q T}
\label{counts_eq}
\end{equation}
\begin{equation}
\widehat{\Phi}_{PF} = \frac{N_T}{q (T - N_T \tau_d)}
\label{pf_eq}
\end{equation}
\begin{equation}
\widehat{\Phi}_{IP} = \frac{1}{q} \frac{N_T - 1}{X_{N_T} - X_1 - (N_T - 1) \tau_d}.
\label{ip_eq}
\end{equation}
\end{subequations}
The first two flux estimators $\widehat{\Phi}_{C}$ (``counts") and $\widehat{\Phi}_{PF}$ (``passive free-running") 
rely purely on the measured photon counts ($N_T$).
Note that \eqref{counts_eq} only counts the number of detected photons during on integration time $T$, normalized by the photon detection probability $q$, while \eqref{pf_eq} also takes into account the dead time ($\tau_d$) of the SPAD after each detection, reducing the effective (active) integration time \cite{pf_spad}. 
Last, $\widehat{\Phi}_{IP}$ (``inter-photon") operates by a slightly different principle, 
since \eqref{ip_eq} takes into account the time between the first and last detected photons ($X_{N_T} - X_1$) within the integration time.
Similarly to \eqref{pf_eq}, the effective integration time is adjusted by the dead time, 
taking into account only the time intervals between the photon arrivals when the SPAD is actually active. For instance, a photon at the very end of the integration time can, in principle, skew the PF-estimator as the dead time beyond end of the integration time may be subtracted.
The reconstruction $\widehat{\Phi}_{IP}$ can be effective in particular at extremely high flux.
Notice that the flux estimators are in units of photons per second on the active area of the SPAD pixel, with $q$ referring to PDP.

The images in TR-MNIST-rec dataset, processed via \eqref{counts_eq}--\eqref{ip_eq}, have been normalized with respect to the median value of the non-zero elements over the whole training set. 
This normalization scheme is more robust than min-max normalization against the high variance outliers in flux estimations (in particularly for $\widehat{\Phi}_{IP}$) due to low photon number. 

In case of the STR-MNIST-rec, we adopt the following flux reconstruction formula from \cite{sync_tr_spad_rec}:
\begin{equation}
 \hat{\Phi}_S   = \frac{\sum_{i=1}^M \mathbf{1} (D_i \neq T_f)} { q \sum_{i=1}^M D_i }, 
 \label{eq:sync_trspad_rec}
\end{equation}
where $\mathbf{1} (\cdot)$ is the indicator function, $D_i$ is the timestamp of the first photon arrival during the frame integration time $T_f$. These formulas are applied to the raw data of STR-MNIST to create the STR-MNIST-rec subset.
From~\eqref{eq:sync_trspad_rec}, we see that, when a pixel detects a photon in every frame, the flux estimate is the inverse of the time spent by the pixel waiting for the first photon, discounted by the 
detection probability $q$. 
If no photons are detected during the frame integration time, then $D_i = T_f$.

Finally, for 1-bit QIS-MNIST simulations, we use the following two flux reconstruction formulas:
\begin{subequations}
\begin{equation}
\widehat{C}_{QIS} = \sum_{t=1}^M B_t
\label{eq:qis}
\end{equation}
\begin{equation}
\widehat{L}_{QIS} = - \ln \left( 1 - \frac{\sum_{t=1}^M B_t}{M} \right)M,
\label{eq:qis_lin}
\end{equation}
\end{subequations}
where \eqref{eq:qis} computes the temporal sum of $M$ binary frames $B_t$, where $t$ is the temporal frame index $t=1,..,M$. In \eqref{eq:qis_lin}, the nonlinear response on the binary frames is taken into account \cite{antolovic2015nonuniformity} and the obtained counts are linearized. Using these reconstruction methods, we form the QIS-MNIST-rec subset. In the QIS-MNIST-rec, no normalization is used such that $\widehat{C}_{QIS} \in  [0, M]$. Note that the linearized $\widehat{L}_{QIS}$ may exceed the value $M$ at high flux level.

To conclude, we generate images from the raw photon data from each dataset and create the corresponding subsets containing the flux reconstructions. This is done in order to process the data to suitable form for the experiments in Section~\ref{ssec:experiments}.

\section{Experiments}
\label{ssec:experiments}

In this section, we conduct experiments on digit classification, using models trained on synthetic data from the SPAD-MNIST dataset. We evaluate how different reconstruction methods at (extreme) low-light conditions affect the classification performance. We also perform digit classification using real SPAD data from a commercial SPAD sensor, to demonstrate the effectiveness of our simulator for creating realistic training data. 

\subsection{Classification on synthetic data}

The original MNIST dataset was introduced together with a simple CNN, LeNet \cite{mnist_lenet}, which was effectively trained to classify digit images. In this section, we evaluate the performance of LeNet when trained on the flux reconstructions from the TR-MNIST-rec, STR-MNIST-rec and QIS-MNIST-rec datasets. 
Although many sophisticated network architectures have been presented, this simple architecture represents a reference for single-photon-based classification in the literature and achieves excellent performance on these datasets.

We trained LeNet on the original MNIST data for 10 epochs using batch size of 64 and Adam optimizer with learning rate of 0.001. LeNet achieved accuracy of 99.22\% using these training parameters. The data was normalized using the normalization scheme described in section~\ref{ssec:recon}. We then trained the same network architecture with the flux reconstructions and frame-based modalities from the TR-MNIST-rec, QIS-MNIST-rec and STR-MNIST-rec datasets using same training parameters. The models were trained separately for each lux level and flux reconstruction pair, in order to investigate the impact of illumination level to the classification accuracy.
Notice that all the flux reconstructions are created from the same initial photon streams, and possible differences may arise from the way that the different SPAD modalities and reconstruction methods process the photon streams.

\begin{figure}[!t]
    \centering
    \includegraphics[width=\columnwidth]{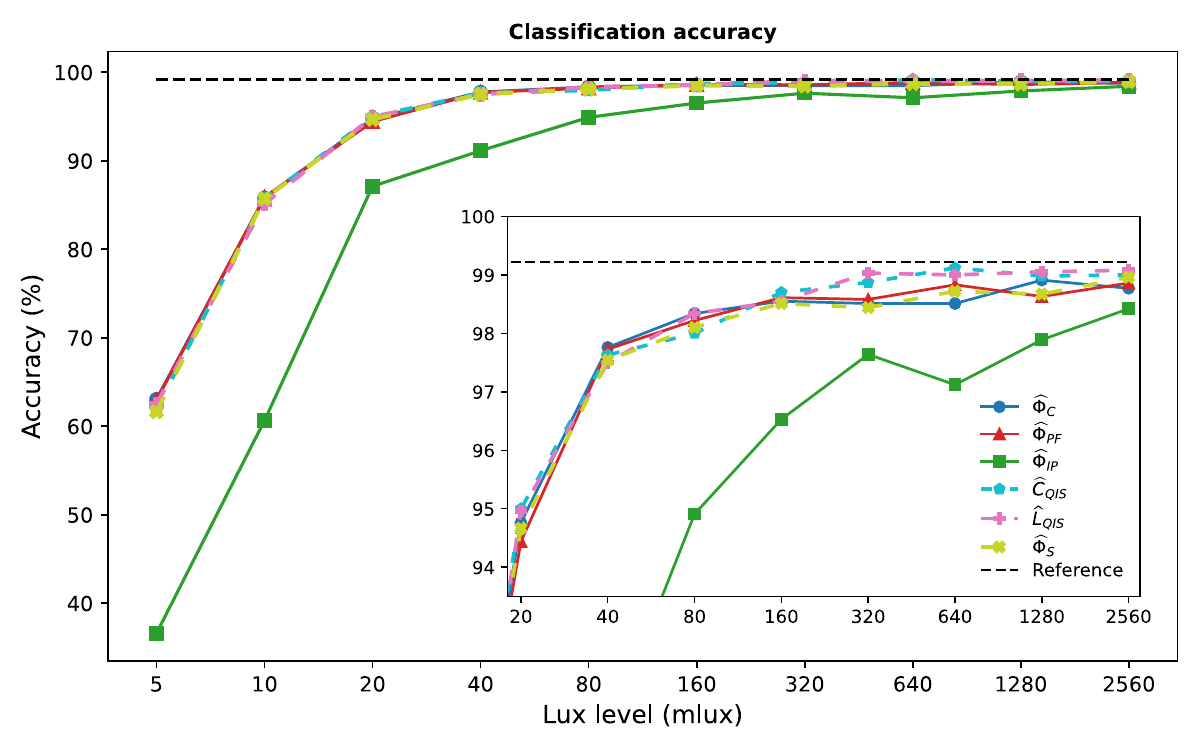}
    \caption{Classification accuracy using different reconstruction methods from the TR-MNIST-rec, QIS-MNIST and STR-MNIST datasets under varying lux levels. The inset shows a zoom-in image for the high accuracy results.}
    \label{fig:accu}
\end{figure}

Fig.~\ref{fig:accu} shows the classification accuracy on all the reconstruction methods (three for TR-MNIST-rec, two for QIS-MNIST-rec and one for STR-MNIST-rec)
at different lux levels.
The inset shows a zoom-in image for the high accuracy results. The black dashed line shows the reference accuracy on the original MNIST data (99.2\%). All reconstruction methods, both from asynchronous photon streams and frame-based reconstructions, yield a similar classification accuracy, except for the IP-SPAD ($\widehat{\Phi}_{IP}$) which struggles at extreme low lux levels. 
Such low performance at low-light is due to the inter-photon time following an exponential distribution, as estimating the flux (proportional to the inverse of the mean inter-photon time) from only a few photon detections can lead to very large variance.
Even at higher lux levels the IP-SPAD accuracy is lower than others due to spurious dark counts in the dark areas of the images that may result in high flux estimates.
While the STR-SPAD data ($\hat{\Phi}_S$) may suffer from a similar type of noise, 
the time-of-darkness waiting for the first photon arrival is averaged over multiple frames, reducing the variance of the flux estimator.
The other reconstruction methods ($\widehat{\Phi}_{C}$, $\widehat{\Phi}_{PF}$, $\widehat{C}_{QIS}$, $\widehat{L}_{QIS}$) are photon count based, and at such low photon counts they all yield very similar results. The effect of dead time or quantization has a low impact on the photon streams and accordingly on classification results. All the reconstruction methods other than IP-SPAD are closely overlapping in the accuracy curves of Fig.~\ref{fig:accu}.

\subsection{Experiments on real data}

To demonstrate that photon data from our simulator can be used to train neural networks for real-world applications, we performed digit classification with real SPAD data. We trained LeNet with the QIS-MNIST dataset using the training parameters described in the previous section. We trained two models, one using simulated data from a single lux level (2560 mlux) and another with data from all lux levels in the dataset (5 to 2560 mlux). We refer to these models as \textit{LUX-2560} and \textit{LUX-ALL}, respectively. This enables us to explore how effectively a model trained on data from one illumination level generalizes to other illumination levels.

\begin{figure}[!ht]
    \centering
    \includegraphics[width=\columnwidth]{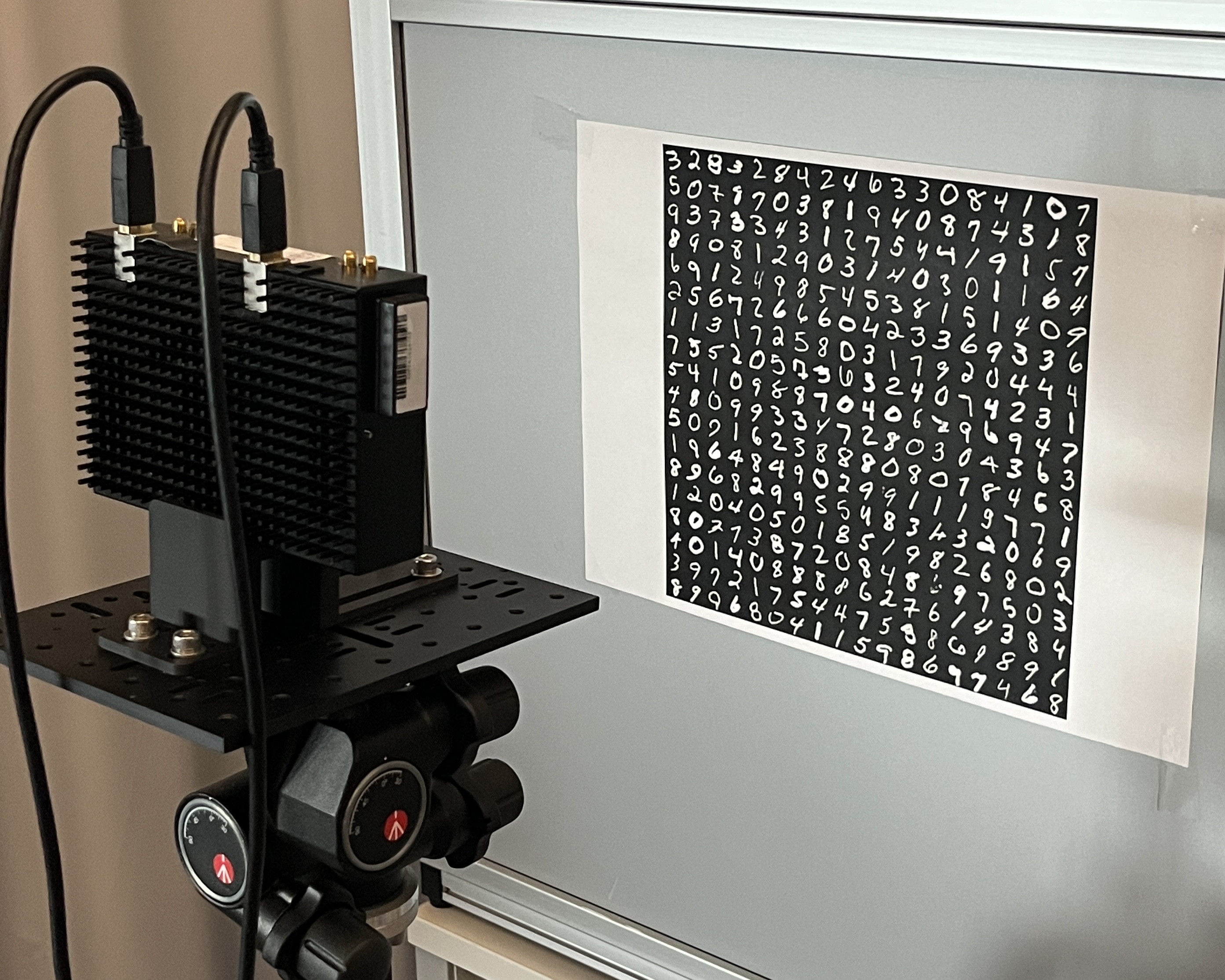}
    \caption{Printed MNIST chart was captured using SPAD512$^2$.} 
    \label{fig:exp}
\end{figure}

\begin{figure*}[!ht]
    \centering
    \includegraphics[width=\textwidth]{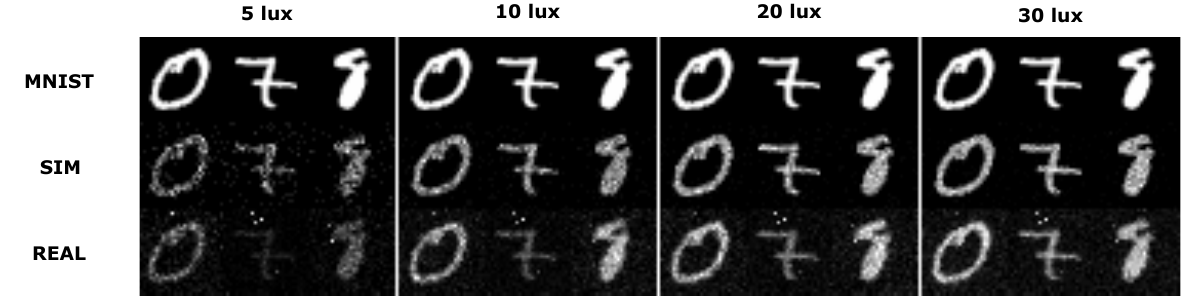}
    \caption{Samples from different lux levels of original MNIST, simulated SPAD data and real images from SPAD512$^2$.} 
    \label{fig:mnist}
\end{figure*}

We randomly selected 256 images from the test set of the original MNIST and printed them into single chart. This chart was then captured using the SPAD512$^2$ camera, with the experimental setup shown in Fig. \ref{fig:exp}. Samples from original MNIST and their corresponding simulated and captured SPAD images at different lux levels are shown in Fig.~\ref{fig:mnist} on respective rows. Both simulated and captured binary frames were processed using \eqref{eq:qis} reconstruction method. For visualization, the images are normalized by setting the maximum pixel value in each image to 1. The normalization illustrates well the impact of hot pixels in the real SPAD data, as their photon detection rate is much higher than normal pixels.  

\begin{table}[t]
    \centering
    \caption{Prediction accuracy on real SPAD data. Accuracy (LUX-2560) is for the network trained merely on the synthetic 2560~mlux data. Accuracy (LUX-ALL) is for a network trained on all lux levels. The accuracy on this subset of the original MNIST data was 96.6\%.}
    \vspace{2mm}
    \begin{tabular}{c c c c}
        Lux level & Lux level & Accuracy & Accuracy \\
        (ambient) & (effective) & (LUX-2560) & (LUX-ALL) \\\hline
        5 lux & 80 mlux & 83.9 \% & 90.9 \% \\
        10 lux & 300 mlux & 90.2 \% & 90.9 \% \\
        20 lux & 500 mlux & 91.0 \% & 91.5 \% \\
        30 lux & 650 mlux & 91.8 \% & 91.8 \% \\
    \end{tabular}
    
    \label{tab:expt}
\end{table}

We repeated the experiment in multiple ambient low-light levels: 5, 10, 20 and 30 lux.
The ambient lux was measured by placing a lux meter at the scene in front of the printed chart.
The simulator expects the lux value on the sensor level; however, in reality only a portion of the light from the ambient illumination actually reaches the sensor (due to optics and other light attenuation factors). Thus, the ambient lux level is not accurate for the simulator lux value setting, and in this section we refer to the flux that reaches sensor as \emph{effective flux}. We estimate the effective lux level from the detected photon counts on the real sensor by comparison with the simulated photon counts (see Table~\ref{tab:expt}).

For each lux level we captured the chart for $K=3$ measurements, with $M = 100$ binary frames each and with 10~$\mu$s integration time, yielding raw data $B_k \in \{0,1\}^{H \times W \times M}$. The frames were then summed together to form an image, according to \eqref{eq:qis}. 
The reconstructed flux images were fed to the LeNet without any pre-processing nor normalization. From the $K$ measurements for each lux level we calculated the mean classification accuracy of \textit{LUX-2560} and \textit{LUX-ALL} and the results are shown in Table \ref{tab:expt}.
As expected, lowest lux data (5~lux) resulted in lowest accuracy of 90.9\% and as the lux level increased, also the accuracy increased. Highest accuracy of 91.8\% was obtained in 30 lux. Out of the two models, \textit{LUX-2560} and \textit{LUX-ALL}, the latter achieved higher accuracy in all illumination levels, confirming it can generalize better to different illumination settings. 

\section{Conclusion}

We presented a realistic SPAD simulator 
that accurately models the behavior and characteristics of SPADs, taking into account the non-uniformities and noise sources which are typical for these sensors. 
We described the simulation pipeline in detail, to enable anyone to re-implement the code easily.
While we present the simulation pipeline for a grayscale images, the simulator can be easily applied to datasets with colors as well, by following the same pipeline in a per-channel manner.

We released the first large SPAD dataset for image classification. We hope that the dataset will encourage the research community to study the SPAD data for different vision tasks and work as a bridge towards SPAD-based imaging for people new to the topic. 
The dataset consists of data for different passive SPAD imaging modalities, as well as, data for various low-light levels with realistic noise sources.

We conducted real world experiments to 1) validate the simulator against real sensors, and 2) test the performance of the LeNet at different light levels.
We performed comprehensive experimental validation of simulation pipeline for the most common passive SPAD modalities, and we demonstrated that synthetic data from the SPAD simulator can be used for training a classification model that can be applied to real SPAD data under different illumination levels.

\bibliographystyle{splncs04}
\bibliography{main}

\end{document}